\documentclass[preprint]{aastex}
\usepackage{graphicx, subfigure}
\graphicspath{{figures/}}
\usepackage{amssymb}
\usepackage{amsmath}
\usepackage{epstopdf}
\usepackage{color}
\usepackage{placeins}
\usepackage{natbib}
\usepackage{enumerate}
\usepackage[retainorgcmds]{IEEEtrantools}
\usepackage{textcomp}
\usepackage{footmisc}
\usepackage{xspace}

\newcommand{\Fermi}{\emph{Fermi}\xspace}

\newcommand{\Swift}{\emph{Swift}\xspace}

\newcommand{\GWNAME}{GW150914\xspace} 
\usepackage{apjfonts} 
\usepackage{amsmath,amstext,natbib}
\usepackage[breaklinks,colorlinks,citecolor=blue,linkcolor=magenta]{hyperref} 
\usepackage{lineno}
\shorttitle{\Fermi-LAT Observations of the LIGO event \GWNAME}
\shortauthors{\Fermi-LAT Collaboration}

\begin{document}

\title{\Fermi-LAT Observations of the LIGO Event \GWNAME}
\author{
M.~Ackermann\altaffilmark{1}, 
M.~Ajello\altaffilmark{2}, 
A.~Albert\altaffilmark{3}, 
B.~Anderson\altaffilmark{4,5}, 
M.~Arimoto\altaffilmark{6}, 
W.~B.~Atwood\altaffilmark{7}, 
M.~Axelsson\altaffilmark{8,9}, 
L.~Baldini\altaffilmark{10,3}, 
J.~Ballet\altaffilmark{11}, 
G.~Barbiellini\altaffilmark{12,13}
% ...
}
\altaffiltext{1}{Deutsches Elektronen Synchrotron DESY, D-15738 Zeuthen, Germany}
\altaffiltext{2}{Department of Physics and Astronomy, Clemson University, Kinard Lab of Physics, Clemson, SC 29634-0978, USA}
\altaffiltext{3}{W. W. Hansen Experimental Physics Laboratory, Kavli Institute for Particle Astrophysics and Cosmology, Department of Physics and SLAC National Accelerator Laboratory, Stanford University, Stanford, CA 94305, USA}
\altaffiltext{4}{Department of Physics, Stockholm University, AlbaNova, SE-106 91 Stockholm, Sweden}
\altaffiltext{5}{The Oskar Klein Centre for Cosmoparticle Physics, AlbaNova, SE-106 91 Stockholm, Sweden}
\altaffiltext{6}{Department of Physics, Tokyo Institute of Technology, Meguro City, Tokyo 152-8551, Japan}
\altaffiltext{7}{Santa Cruz Institute for Particle Physics, Department of Physics and Department of Astronomy and Astrophysics, University of California at Santa Cruz, Santa Cruz, CA 95064, USA}
\altaffiltext{8}{Department of Physics, KTH Royal Institute of Technology, AlbaNova, SE-106 91 Stockholm, Sweden}
\altaffiltext{9}{Tokyo Metropolitan University, Department of Physics, Minami-osawa 1-1, Hachioji, Tokyo 192-0397, Japan}
\altaffiltext{10}{Universit\`a di Pisa and Istituto Nazionale di Fisica Nucleare, Sezione di Pisa I-56127 Pisa, Italy}
\altaffiltext{11}{Laboratoire AIM, CEA-IRFU/CNRS/Universit\'e Paris Diderot, Service d'Astrophysique, CEA Saclay, F-91191 Gif sur Yvette, France}
\altaffiltext{12}{Istituto Nazionale di Fisica Nucleare, Sezione di Trieste, I-34127 Trieste, Italy}
\altaffiltext{13}{Dipartimento di Fisica, Universit\`a di Trieste, I-34127 Trieste, Italy}
% ...

\clearpage

\begin{abstract}
The \Fermi Large Area Telescope (LAT) has an instantaneous field of view covering $\sim 1/5$ of the sky and completes a survey of the entire sky in high-energy gamma rays every ~3 hours. It enables searches for transient phenomena over timescales from milliseconds to years. Among these phenomena could be electromagnetic counterparts to gravitational wave sources. In this paper, we present a detailed study of the LAT observations relevant to  Laser Interferometer Gravitational-wave Observatory (LIGO) event \GWNAME~\citep{LIGO_GW150914}, which is the first direct detection of gravitational waves and has been interpreted as due to the coalescence of two stellar-mass black holes.
The localization region for \GWNAME was outside the LAT field of view at the time of the gravitational-wave signal.
However, as part of routine survey observations, the LAT observed the entire LIGO localization region within $\sim 70$ minutes of the trigger, and thus 
enabled a comprehensive search for a $\gamma$-ray counterpart to \GWNAME. The study of the LAT data presented here did not find any potential counterparts to \GWNAME, but it did provide limits on the presence of a transient counterpart above 100 MeV on timescales of hours to days over the entire \GWNAME localization region. 
\end{abstract}

\maketitle

\section{Introduction}


The $\sim$2.4 sr field of view (FoV) and 
broad energy coverage from 20 MeV to $>$300 GeV of the Large Area Telescope (LAT, \citealt{2009ApJ...697.1071A}) on the {\it Fermi Gamma-ray Space Telescope} mission make it a powerful instrument to monitor the sky for high-energy transients.  As the LAT surveys the entire sky every 3 hours, it is sensitive to transient emission from a variety of sources, including stellar-mass compact objects (Neutron Stars -- NS, and Black Holes -- BH) over timescales from milliseconds to years, including those predicted to be associated with gravitational waves (GWs). Current GW detectors are sensitive to signals from the merging of compact objects. Some of these mergers, like the putative progenitors of short Gamma-Ray Bursts (sGRBs, \citealt{1986ApJ...308L..43P,1989Natur.340..126E,1992ApJ...395L..83N,2011ApJ...732L...6R,2012ApJ...746...48M}), emit both a short-lived $\gamma$-ray signal ($\lesssim 2$ s) immediately after the merger (``prompt'' emission), and a long-lived and broadband ``afterglow'' signal lasting minutes to hours. If the GRB happens to be in the FoV at the time of the trigger, the LAT can detect the short-lived prompt emission phase.
If the GRB is outside the FoV, because of its survey capability the LAT can still detect the GRB by measuring its temporally-extended afterglow emission as soon as it enters the FoV. The LAT has detected high-energy $\gamma$-ray emission from $>$130 GRBs to-date \citep{2016AAS...22741601V}, including $\sim$10 sGRBs. Given the uncertainty in theoretical predictions for counterparts to GW sources and the demonstrated emission of high-energy $\gamma$-rays from systems containing compact objects, searching the LAT data for $\gamma$-ray counterparts to GW events.


The era of gravitational wave astronomy has begun with the first science run (`O1') of the recently upgraded 
LIGO \citep{1992Sci...256..325A,2009PhRvD..80j2001A} from 2015 September to 2016 January. The Virgo Observatory \citep{1999APh....10..369C, 2009CQGra..26h5009A} will soon be added to the network for the second science run in late 2016. The GW frequency range that LIGO and Virgo are sensitive to is expected to be dominated by mergers of compact stellar-mass objects that are most likely remnants of 
stellar evolution: two neutron stars (NS-NS), two black holes (BH-BH), or a NS and a BH. The sensitivity and horizon distance of the GW network to these mergers scales with the masses of the systems; therefore the accessible volume of the Universe for NS mergers is significantly smaller than that of BH mergers.  
When those mergers include at least one NS, an electromagnetic counterpart is predicted to accompany the merger signal in the form of a sGRB. The electromagnetic outcome of a BH-BH merger is less well understood. Finding the counterpart of a GW event is important for understanding the nature of the source. It also has an additional yet less-evident benefit: it improves significantly the accuracy with which all parameters (distance, mass, spin, inclination, etc.) can be estimated. This is obtained by better constraining the localization of the event, which is normally poorly estimated using only the GW signal \citep{2009LRR....12....2S}. \Fermi-LAT can localize a transient source with sub-degree accuracy, a very big improvement with respect to a typical localization region from a GW detector, which will typically cover hundreds of square degrees.

On 2015 September 14 at 09:50:45 UTC the LIGO Hanford and Livingston installations  detected a coincident signal, within 10 ms, from \GWNAME, a high-significance trigger in the engineering run just prior to the start of O1. The trigger was determined to be consistent with a waveform predicted by General Relativity from the inspiral and merger of a stellar-mass binary BH system, with constituent BHs of masses around 29 M$_\odot$ and 36 M$_\odot$ \citep{LIGO_GW150914}. The GW luminosity expected theoretically for a massive BH-BH merger leads to an estimate of around 400 Mpc (i.e., z$\sim 0.09$, \citealt{2041-8205-818-2-L22}) for the distance of the source. This observation provides evidence for the existence of isolated and binary stellar-mass black holes, and the first observation of such a system merging. Two days later, on 2015 September 16, LIGO notified the electromagnetic (EM) partner observatories operating within a Memorandum of Understanding (MOU). The EM partner observatories executed follow-up programs \citep{LVEM} to search for a counterpart within the 750 deg$^2$ localization region ($\sim 90\%$ confidence), which was later refined to 601 deg$^2$ \citep{LVEM}. \Fermi was operating in normal survey mode at the time of the trigger. Hence, the LAT autonomously observed the entire LIGO localization region within $\sim 70$ minutes of the GW trigger, independently of any notification from LIGO, in the high-energy $\gamma$-ray band. 
The LAT Collaboration reported a preliminary search throughout the LIGO localization area that did not reveal any new $\gamma$-ray sources \citep{2015GCN..18709...1O}. The results of a search of the data of the other instrument on board \Fermi, the Gamma-Ray Burst Monitor \citep[GBM,][]{GBMinstrument} and the evidence for a weak counterpart are discussed separately in \cite{GBMLIGO}.

In this paper, we describe LAT observations of the localization area of \GWNAME around the time of the trigger, including a dedicated search for an EM $\gamma$-ray counterpart.  No candidate counterparts were found. We describe the details of the data analysis in  \S\ref{sec:data_analysis}, discuss the implications of the counterpart search and prospects for future GW triggers in \S\ref{sec:discussion}, and conclude in \S\ref{sec:conc}.

\section{Data analysis} \label{sec:data_analysis}

We performed two complementary sets of searches for transient high-energy $\gamma$-ray emission:  automated searches (\S\ref{sec:auto_search}) that are performed routinely on all LAT data, and targeted searches in the LIGO localization region (\S\ref{sec:ligo_cont}) on short and long time baselines that exploit the full sensitivity of the standard LAT analysis chain. In Appendix~\ref{apex:likelihood}, and \ref{sec_tsmaps} we provide more details on the various analysis steps.


\subsection{Automated Searches}\label{sec:auto_search}

Since the launch of \Fermi in 2008, automated on-board and on-ground analyses of GBM and LAT data have been in place to search for new transients at various time- and energy-scales. As our understanding of the instruments, data, and the variable and transient $\gamma$-ray sky has improved, so have our automated analyses.
Three LAT pipelines were relevant to the counterpart search for \GWNAME:
\begin{itemize}
\item The Burst Advocate (BA) Tool and LAT Transient Factory (LTF) are automated pipelines that search for excess emission in the LAT data at the positions of triggers from GBM, \Swift, INTEGRAL and MAXI at the time of the trigger and intervals in the hours afterwards.  As there were no on-board triggers by any of these instruments coincident with \GWNAME, the BA Tool and LTF were not initiated.  However, in the event of a LAT on-board trigger or a trigger from these observatories coincident with a GW trigger, the pipelines would perform an automated search once the LAT data were available on the ground ($\sim 8$ hours after trigger), with results monitored by on-shift personnel.
\item Automated Science Processing (ASP; \citealt{ASP}) is the standard LAT search for transient or flaring sources on 6-hour, 24-hour, and 7-day timescales. The ASP pipeline performs a detection step via a blind search for sources on all-sky counts maps constructed from the event data acquired at each timescale; then, a standard likelihood analysis is run on those data using a source model that includes the candidate sources found in the detection step as well as the already known sources that have been designated for regular monitoring.
LAT Flare Advocates (a.k.a. Gamma-ray Sky Watchers) offer a prompt human verification service to the automatic pipelines and review the results daily, providing an internal report to the LAT Collaboration.
Relevant information on monitored, flaring and transient sources is released to the astrophysical community using the LAT multiwavelength mailing-list\footnote{To sign up for the LAT multiwavelength list visit \url{http://fermi.gsfc.nasa.gov/ssc/library/newsletter/}}, Astronomer's Telegrams\footnote{\url{https://www-glast.stanford.edu/cgi-bin/pub_rapid}} and Gamma-ray Coordinates Network notices\footnote{\url{http://gcn.gsfc.nasa.gov/fermi_lat_mon_trans.html}}. Weekly summary digests are made available through the \Fermi Sky Blog\footnote{\url{http://fermisky.blogspot.com}}.
The LAT Flare Advocate service has been very effective in identifying potential candidates for quick follow-up and coordinated observations at other wavelengths \citep{2013arXiv1303.4054C}. 
ASP discovers an average of 8 previously unknown $\gamma$-ray transients per year, and has also detected bright GRB afterglows \citep[e.g., GRB 130427A,][]{2014Sci...343...42A}.
%
\item \Fermi All-sky Variability Analysis (FAVA) is a photometric technique that searches for new transients and variable sources on 7-day timescales \citep{2013ApJ...771...57A}. This method compares the number of detected $\gamma$ rays with the average number of expected  $\gamma$ rays based on the observed long-term average in a given region of the sky. In this way, FAVA provides a computationally inexpensive blind search of flux variations over the entire sky that is independent of both an assumed spectral shape of the flaring source and any model for the diffuse $\gamma$-ray background. The FAVA pipeline detects an average of 16 flares per week; about 10\% of these are not associated with $\gamma$-ray catalog sources \citep[e.g.,][]{2014ATel.6098....1K}. Seven-day FAVA lightcurves for any position on the sky are publicly available at the FAVA Data Portal\footnote{\url{http://fermi.gsfc.nasa.gov/ssc/data/access/lat/FAVA/}} hosted at NASA's \Fermi Science Support Center (FSSC).

\end{itemize}
During the 6-hour interval\footnote{The ASP 6-hour interval containing the LIGO trigger time includes LAT data between 2015 September 14, 06:11:33$-$12:00:00 UTC.} containing the LIGO trigger \GWNAME, ASP detected ($>$3$\sigma$) twelve known $\gamma$-ray sources and three low-significance ($>$1$\sigma$) unidentified transients, none consistent with the LIGO event localization. None of the LAT pipelines found a possible counterpart to \GWNAME. 
%

\clearpage

\subsection{Search in the LIGO localization region}\label{sec:ligo_cont}
The LIGO Scientific Collaboration reported results from a Bayesian parameter estimation analysis of \GWNAME under the assumption that the signal arises from a compact binary coalescence (CBC) using the latest offline calibration of the GW strain data. 
The most accurate localization map for this event (\texttt{LALInference}) is 
based on a Bayesian Markov-Chain Monte Carlo and nested sampling to forward model the full GW signal including spin precession and regression of systematic calibration errors. 
The localization probability is primarily in the southern portion of the annulus determined by the arrival time difference between LIGO Hanford and LIGO Livingston of $\sim 7$ ms.

Given the uncertainty on EM signals from the merging of two BHs, 
we searched different time windows by carrying out two customized analyses of the LAT data. Both analyses are based on the standard maximum likelihood analysis technique used for LAT data, and summarized in Appendix~\ref{apex:likelihood}. In all our searches we included in the likelihood model all sources (point-like and extended) from the LAT source catalog \citep[3FGL,][]{2015ApJS..218...23A}, as well as the Galactic and isotropic diffuse templates provided by the \Fermi-LAT collaboration\footnote{\url{http://fermi.gsfc.nasa.gov/ssc/data/access/lat/BackgroundModels.html}}. We used the Pass~8 \texttt{P8\_TRANSIENTR010E\_V6} event class and the corresponding instrument response functions. These searches are described in the following sections.\\  
 
\subsubsection{Short-baseline search}\label{sec:short_search}

\begin{figure*}[t]
\begin{center}
  \includegraphics[width=\textwidth]{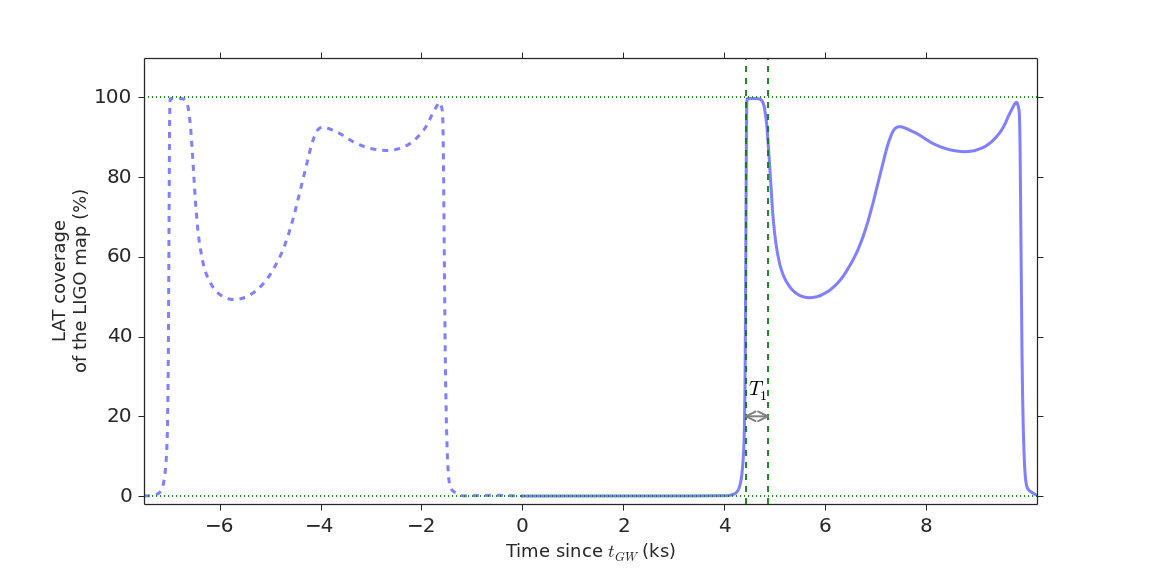}
\caption{\Fermi-LAT coverage (see text) of the LIGO localization map, as a function of time, before $t_{gw}$ (dashed curve) and after $t_{gw}$ (solid curve).
The green dashed vertical lines denote the boundary of the interval $T_{1}$ used in the analysis (see text for details), while the horizontal dotted lines mark respectively 0\% and 100\% coverage.}
\label{fig:coverage}
\end{center}
\end{figure*}

This search focuses on the hours immediately after the GW trigger $t_{GW}$ (2015 September 14, 09:50:45 UTC). The LAT can detect long and short GRB afterglows up to thousands of seconds after the end of the prompt emission \citep{2010ApJ...709L.146D,2013ApJS..209...11A, 2015arXiv150203122V}. Thus a search in this time window is the most likely to find a counterpart to \GWNAME if it is similar to a sGRB.
\Fermi was in normal survey mode operations around $t_{GW}$, rocked 50$^\circ$ North from the orbital plane. We consider a point in the sky observable by LAT if it is within the 65$^\circ$ radius FoV and has an angle with respect to the local zenith smaller than 100$^\circ$. The latter requirement is used to exclude contamination from $\gamma$-ray emission from the Earth's atmosphere. The {\it coverage} is the integral of the probability densities of all points in the LIGO localization probability map observable by LAT at a given time, and it is shown in Fig.~\ref{fig:coverage} as a function of time.
While the coverage was between 50\% and 90\% in the hours before the trigger (dashed line), at $t_{GW}$ the LAT was unfortunately viewing  the opposite side of the sky from the LIGO localization region. The coverage was zero until $\sim t_{GW} + 4200$. The time interval $t_{GW}+4442$--$4867$\,s ($T_1$) had coverage $>$90\%, while during $t_{GW}+4867$--$10000$\,s coverage varied between 50 and 98\%, decreasing back to zero at around $t_{GW} + 10$ ks. We searched for a transient source in the time interval having more than 90\% coverage ($T_1$), and we did not find any significant excess. 

\begin{figure*}[tb]
\begin{center}
  \includegraphics[width=0.45\textwidth]{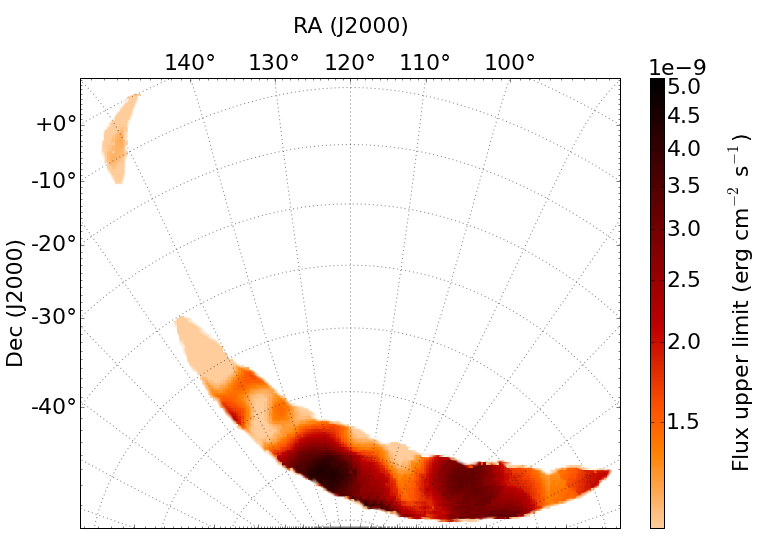}
  \includegraphics[width=0.45\textwidth]{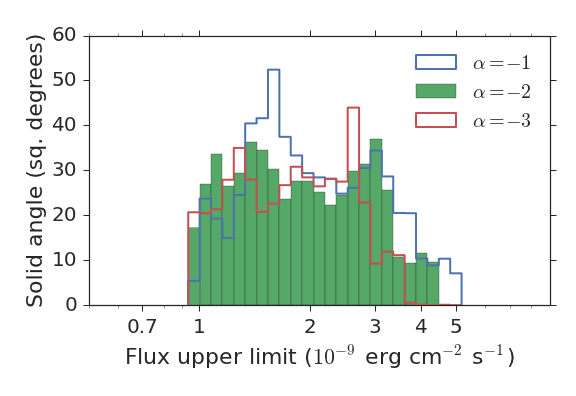}
\caption{Flux upper limits (95\% c.l.) in the energy range 100 MeV--1 GeV for \GWNAME during the interval $T_{1}$ (4442--4867 s from $t_{GW}$). Left panel: upper limits map covering the 90\% region of the LIGO probability map. Right panel: histogram of the upper limits in the map. We assumed a power-law spectrum for the source, with a photon index $\alpha$ of $-2$ (typical of afterglows of GRBs, green histogram), $-1$ (blue histogram) and $-3$ (red histogram). While the distributions are slightly different for the three cases, the ranges spanned by the upper limits are largely independent of the photon index. }
\label{fig:ul_map}
\end{center}
\end{figure*}

We then derived upper limits for the $\gamma$-ray flux of \GWNAME.
Since the sensitivity of the LAT depends strongly on the angle from the source to the boresight of the instrument, the continuous variation of the LAT viewing direction in survey mode makes any flux limit for a particular source position time-dependent.  Flux limits are also sensitive to astrophysical backgrounds, particularly in the Galactic plane, so that positions along the LIGO arc will have different flux limits, even for the same observing conditions.  These effects mean that flux limits vary according to both time of observation and position in the localization region.
We show a map of the derived upper limits (95\% confidence level) for the $\gamma$-ray flux of GW150914 in the band 100 MeV $-$ 1 GeV in the left-hand panel of Fig.~\ref{fig:ul_map}, and a histogram of the upper limits in the right-hand panel, both for interval $T_{1}$. Assuming a power-law spectrum for the source with a photon
index of $\alpha = -2$, which is typical for GRB afterglows at LAT
energies, the upper limits we find have a median of $1.7 \times
10^{-9}$ erg cm$^{-2}$s$^{-1}$, and 5\% and 95\% percentiles of $0.9 \times
10^{-9}$ and $3.7\times 10^{-9}$ erg cm$^{-2}$s$^{-1}$, respectively.
These upper limits are only weakly dependent on the choice of
$\alpha$ as shown in the right-hand panel of Fig. \ref{fig:ul_map}.
We now describe the upper limits analysis in more detail.\\

We considered all $\gamma$-rays with reconstructed energies between 100 MeV and 100 GeV. We then considered the \texttt{LALInference} probability map provided by LIGO, which is a \texttt{HEALPix} map with \texttt{NSIDE = 512} \citep{2005ApJ...622..759G}, corresponding to a typical pixel size of $\sim$$0\fdg11$. The point-spread function (PSF) of the LAT has a 68\% containment radius at 1 GeV of $\sim$1$^\circ$. To save computation time, we resampled the map to \texttt{NSIDE = 256}, which corresponds to a typical pixel size of $\sim$0$\fdg2$. We considered the 11463 pixels in the resampled map contained within the 90\% containment region provided by LIGO. Let us denote $\vec{v}_{i}$ the direction of the center of the i-th pixel. For each i-th pixel we performed an independent likelihood analysis (see Appendix~\ref{sec:auto_search}), considering a Region Of Interest (ROI) of 8$^\circ$ centered in $\vec{v}_{i}$ and testing for a new source at the position $\vec{v}_{i}$. For each likelihood analysis we required the zenith angle of the events to be no more than 100$^\circ$. We did not find any new source with a test statistic (TS) above our adopted threshold of 25, corresponding to $\sim 5 \sigma$ (pre-trials). We then computed the 95\% confidence level upper limit for the flux of a source at each $\vec{v}_{i}$. In order to obtain upper limits reasonably independent of the photon index $\alpha$ adopted in the analysis we need to choose an energy range small enough. We have chosen the range 100 MeV--1 GeV for the upper limits, which provides the largest photon statistic and the maximum sensitivity for sources similar to GRBs. The right-hand panel of Figure~\ref{fig:ul_map} shows that our measurement is indeed largely independent of the choice of $\alpha$.\\

\cite{GBMLIGO} reported the weak transient $\gamma$-ray source \GWNAME-GBM lasting $\sim 1$ s, 0.4 s after the LIGO trigger on \GWNAME. \GWNAME-GBM is consistent with being due to a low-fluence sGRB at an unfavorable viewing geometry to the GBM detectors, although this is not expected from a BH-BH merger.
Assuming the two signals have a common origin, the
combined LIGO and GBM observations reduce the 90\% confidence region from 601 deg$^2$ to 199 deg$^2$. Within the combined LIGO/GBM localization, and in the time interval $T_{1}$, the most significant excess in LAT data has TS = 18. We estimate for this excess a p-value of $\sim 0.05$, taking into account the number of trials of our analysis, which corresponds to a significance that is well below our threshold of $5 \sigma$. The excess has a spectrum well modeled with a power law with a soft photon index $\alpha = -3.2 \pm 0.8$, and it is located close the limb of the Earth (which has indeed a soft spectrum). Therefore we consider this excess in the LAT data very likely to be either a statistical fluctuation or due to Earth limb contamination, and therefore unrelated to \GWNAME or \GWNAME-GBM.


\subsubsection{Long-baseline search}\label{sec:long_search}

\begin{figure*}[h!t!b!]
\begin{center}
  \includegraphics[width=1.0\textwidth,trim={1cm 0 1cm 0},clip]{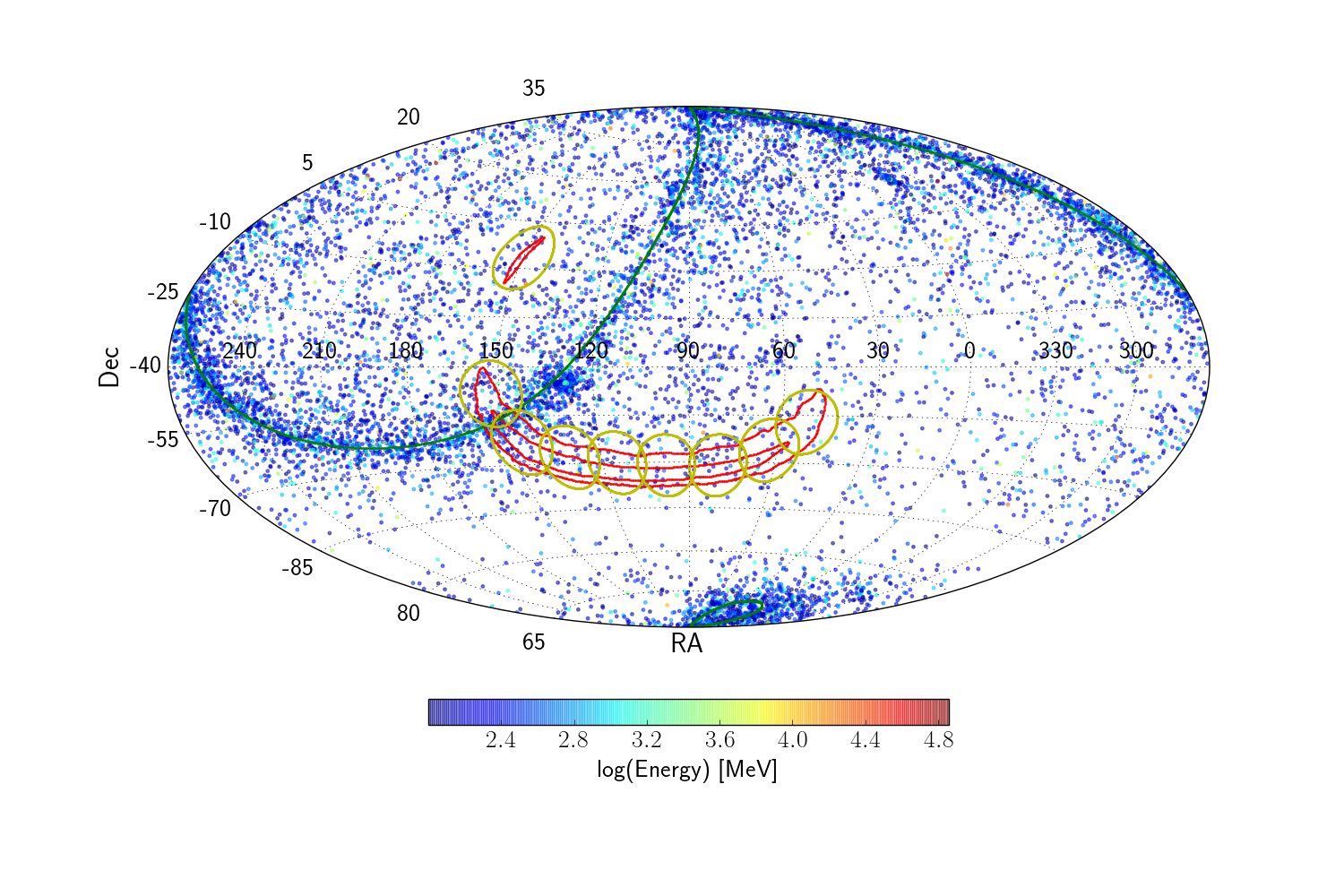}
\caption{LAT-detected $\gamma$-rays in celestial coordinates (J2000) in the interval from $t_{GW}$ to $t_{GW}$+$10000$s, and in the energy range 60 MeV--100 GeV. The dots are colored according to energies of the $\gamma$ rays as indicated in the color bar. The highest-energy photon recorded in this time interval has an energy of $\sim 60$ GeV, corresponding to the maximum of the color bar. The 90\% and 50\% contour levels from \texttt{LALInference} are overlaid on the image, in red. The map is in \texttt{Hammer-Aitoff} projection and is centered at R.A., Dec$=90^{\circ}$, $-40^{\circ}$. The Galactic plane is highlighted in green. The nine LAT ROIs where we perform the searches described in \S\ref{sec:long_search} are shown as yellow circles.}

\label{fig:contours}
\end{center}
\end{figure*}

In this second search we considered data gathered during a 2-month interval centered on $t_{GW}$. In order to increase the number of $\gamma$-rays we included all photons with energies between 60 MeV and 100 GeV. Since the PSF at 60 MeV is broad, we applied a zenith cut of 95$^\circ$ to further limit Earth limb contamination.
We looked both for a long-duration signal of the order of one day, as well as for a short-duration signal but not necessarily in strict temporal coincidence with the LIGO trigger.
To this end, we covered the entire 90\% probability region provided by LIGO with a set of nine partially overlapping ROIs, each with a radius of 10$^\circ$. Figure \ref{fig:contours} shows the locations of these ROIs and the confidence contours obtained from the \texttt{LALInference} probability map. They are overlaid on a sky map of the $\gamma$-rays detected by the LAT over the interval $t_{GW}$ to $t_{GW} + 10000$ s. In Table~\ref{table:roi} we provide the location of the center of each ROI, listing all the LAT sources from the 3FGL catalog \citep{2015ApJS..218...23A} within each ROI, along with their associations and their classes, using the same notation as the 3FGL catalog. 

For the first analysis of the second search we divided the data in 10~ks time bins. For each time bin and for each ROI we calculated a TS map (see Appendix~\ref{sec_tsmaps}) and  determined the location of the grid position with the maximum TS (TS$_{max}$). We considered the position of TS$_{max}$ as the location of a possible counterpart, and we ran an unbinned likelihood analysis adding a point source at the position of TS$_{max}$. This gave a value of TS$_{src}$ (which is normally similar to TS$_{max}$).
In these maps derived from low-statistics data single high-energy $\gamma$-rays can cause a high value of TS$_{max}$. 
To reduce the number of false positives from random coincidences, we required that the number of photons N$_{\gamma}$ that have a probability larger than 0.9 to be associated with the candidate counterpart to be greater than 2.  No excesses met this requirement. In Appendix~\ref{sec_tsmaps} we present the TS maps for the 9 ROIs for the time interval 0--10~ks since $t_{GW}$. We repeated the same analysis considering time bins of 1 day, and again did not find any significant excess.

We also considered the possibility of excesses over shorter timescales ($<$1 h), similar to the typical duration of high-energy emission from GRBs \citep{2013ApJS..209...11A,2015arXiv150203122V}, but not in temporal coincidence with the GW trigger and hence not covered by the ``short-baseline'' search described in \S\ref{sec:short_search}. 
We calculated the entry and exit times for each ROI in the FoV of the LAT (a "FoV passage"), requiring that the distance between the LAT boresight and the center of the ROI be $<$60$^\circ$.
In standard survey mode, the duration of a FoV passage varies from a few hundred seconds to nearly one hour. Since we do not know if an EM signal would be in temporal coincidence with the GW signal, we searched for possible excesses in every passage, corresponding to a total of 6615 passages for each ROI. We did not detect any significant excess in any of the passages before or after $t_{GW}$ for any ROI.


\begin{figure}[!ht]
\begin{center}
  \includegraphics[width=0.45\textwidth]{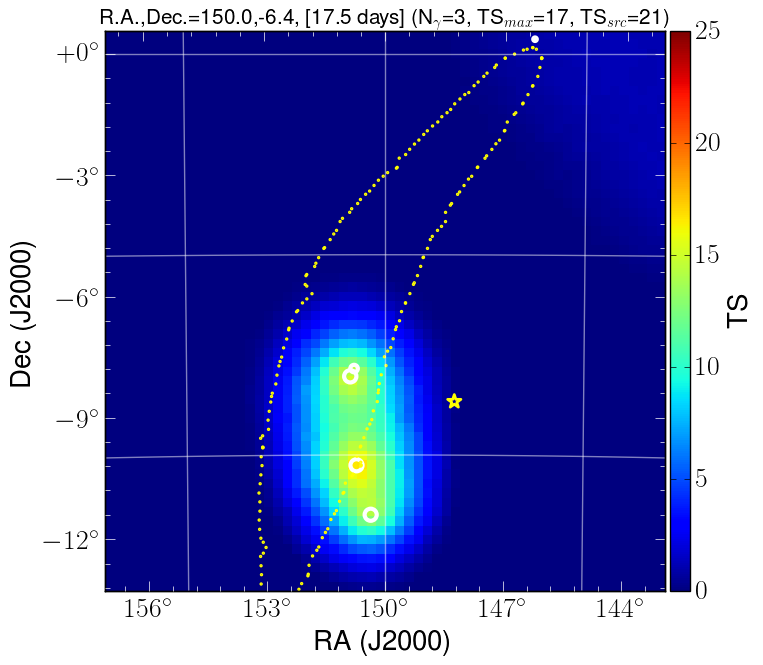}
  \includegraphics[width=0.45\textwidth]{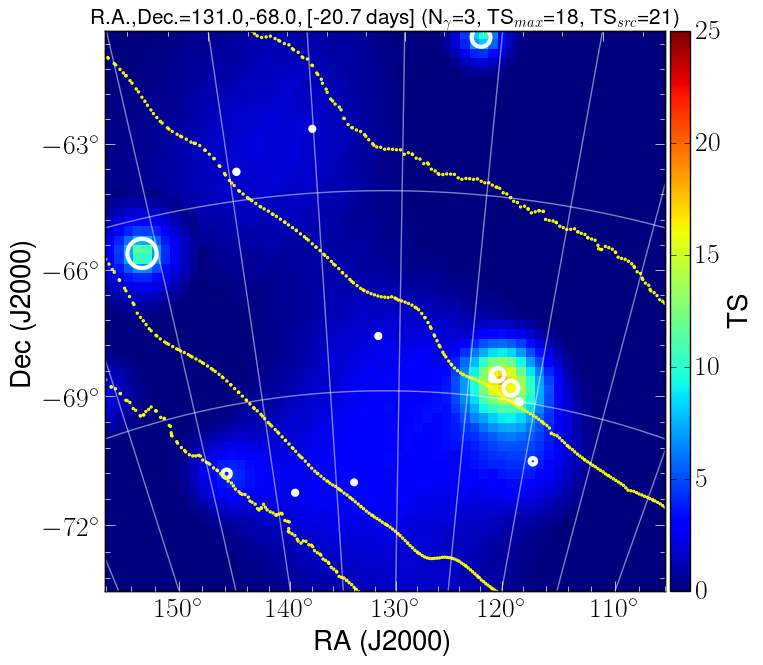}\\
  \includegraphics[width=0.45\textwidth]{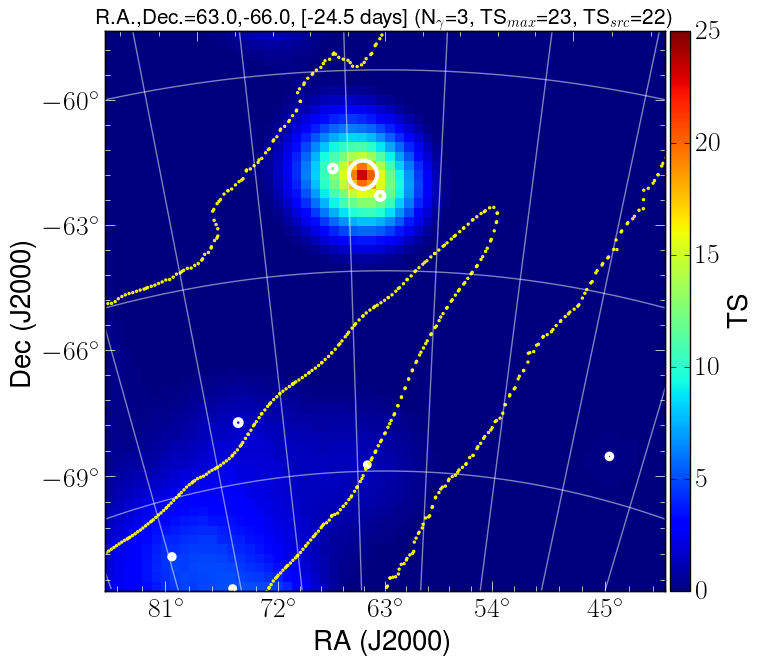}
  \includegraphics[width=0.45\textwidth]{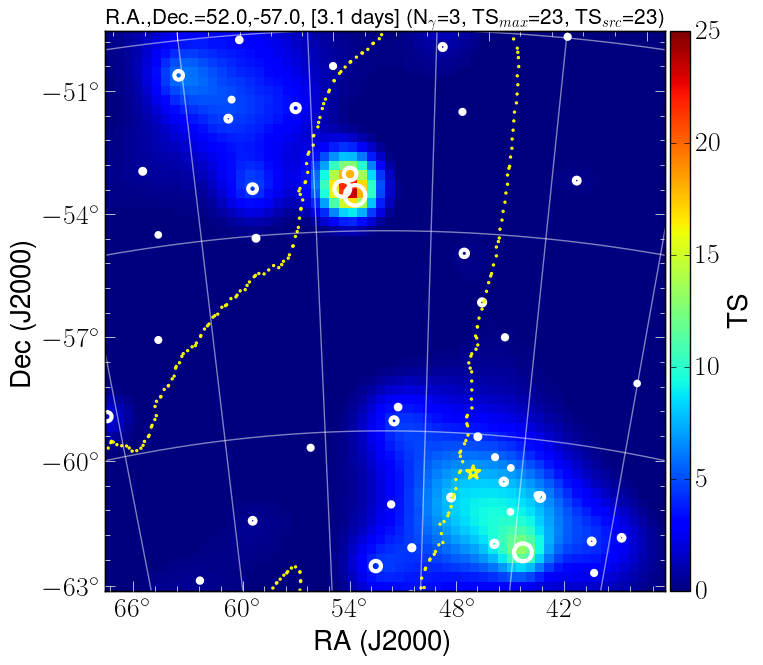}
\caption{TS maps of the most significant excesses detected over one-orbit timescales ($\sim$ minutes) within 30 days before or after the LIGO trigger. Yellow stars are sources from the 3FGL catalog, and circles are individual LAT $\gamma$ rays with their size proportional to the reconstructed energy. The yellow dots trace the LIGO 50\% and 90\% confidence contour.}
\label{fig:tsmap}
\end{center}
\end{figure}

Figure~\ref{fig:tsmap} displays the four TS maps with the highest TS$_{src}$ and N$_{\gamma}>2$.
The first map (TS$_{src}=21$) is from the first ROI and corresponds to the passage from 22:15:53 to 22:20:30 UTC on 2015 October 10 (17.5 days after the trigger time $t_{GW}$).
The second map is from the third ROI in the passage from 17:47:41 to 17:56:41 UTC on 2015 August 21 (20.7 days before $t_{GW}$), and has also a TS$_{src}=21$.
The third map (TS$_{src}=22$) is from the seventh ROI in the interval from 21:35:48 to  21:50:01 UTC on 2015 August 20 (24.5 days before $t_{GW}$).
Finally, the last map is from the eighth ROI, during 12:59:40 to 13:39:41 UTC on 2015 September 17 (3.1 days after the trigger LIGO trigger time $t_{GW}$) has the highest value of TS with TS$_{src}=23$. 
The peak TS values correspond to low-energy $\gamma$ rays in random coincidence with high-energy $\gamma$ rays, and the highest values obtained in this analysis should not be considered indicative of a possible EM counterpart.
Moreover, the above search involves a large number of trials. To estimate their impact on the peak TS values we performed a Monte Carlo study described in the next section.

\subsection{Comparison with Monte Carlo simulations}

To validate our interpretations of TS values we performed a detailed Monte Carlo simulation of 2 months of data (the same interval used in our analysis). 
The actual pointing history of the satellite was used; therefore the correct exposure of the sky was automatically taken into account.
All the sources from the 3FGL catalog were kept fixed at their 3FGL catalog fluxes.
As a result, the simulation is suitable for computing the distribution of TS under the null hypothesis that no transient signal is present.
With the simulated data we repeated exactly the same analysis used on real data and described in the previous section. 

\begin{figure}[ht]
\begin{center}
  \includegraphics[width=0.7\textwidth]{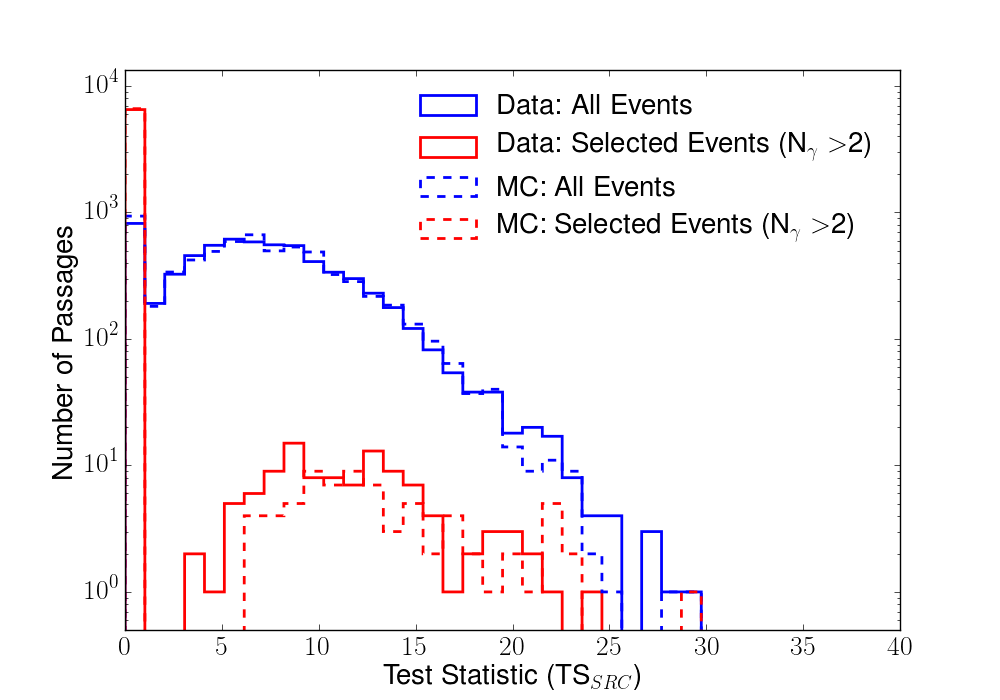}
\caption{Data-Monte Carlo comparison of the distributions of the TS$_{src}$ value. Solid lines correspond to TS distributions obtained from data, while dashed lines correspond to distributions obtained from Monte Carlo simulations. Blue lines are the distributions of the TS values when no additional selections are applied, while red corresponds to the final distribution when we require that at least three $\gamma$-rays have at least a 90\% probability to be associated to the additional point source.}
\label{fig:tsdistribution}
\end{center}
\end{figure}

In Figure \ref{fig:tsdistribution} we compare the distribution of TS$_{src}$ obtained from flight data (filled lines) and Monte Carlo (dashed lines). 
We note that the Monte Carlo distributions are a good match to the distributions of the TS$_{src}$ values obtained from the flight data, and the good absolute agreement is consistent with no statistically-significant transient counterpart being present in the flight data.
Also, given the large number of trials, relatively high values of TS can be obtained in Monte Carlo simulations even if no transient signal was added.
In the flight data we found 4 cases with TS$>$20 and with N$_\gamma > 2$, and this must be compared with the 9 cases we obtain when we analyzed the simulated data. In other words, we expected 9 false positives with TS$_{src}>20$ (and N$_{\gamma}>2$) in 2 months of flight data, and we observe 4.

\section{Discussion} \label{sec:discussion}



The most promising astrophysical GW sources in the frequency range of LIGO/Virgo are the mergers of compact object binaries with NS and/or stellar-mass BH constituents.  The detection of \GWNAME is undoubtedly a major breakthrough in astronomy, being the first detection of GWs and the first detection of a merging binary BH system.  

Maximizing the science return from GW detections requires the identification and study of coincident EM counterparts, which  
would help resolve degeneracies associated with the inferred binary parameters. For example, a strong $\gamma$-ray signal may only be seen at small binary inclination relative to the sky, such that the jet (along the direction of the total angular momentum) is pointed toward us. A high-significance counterpart association would also decrease the significance threshold 
necessary for a confident GW detection, thereby effectively increasing the distance to which the GW signal can be detected and the searchable volume as the cube of the distance.  It would help identify the host galaxy and thereby constrain or measure the merger redshift, 
setting the luminosity scale and allowing an independent measurement of the Hubble constant or other cosmological parameters \citep{2012ApJ...746...48M,2012ApJ...748..136C}.  The complementary information encoded in the EM signal (spectral and temporal evolution, energetics, inferred environment) is likewise essential to unravel the astrophysical context of the coalescence event.

\begin{figure}[t]
\centering
\includegraphics[width=0.7\textwidth]{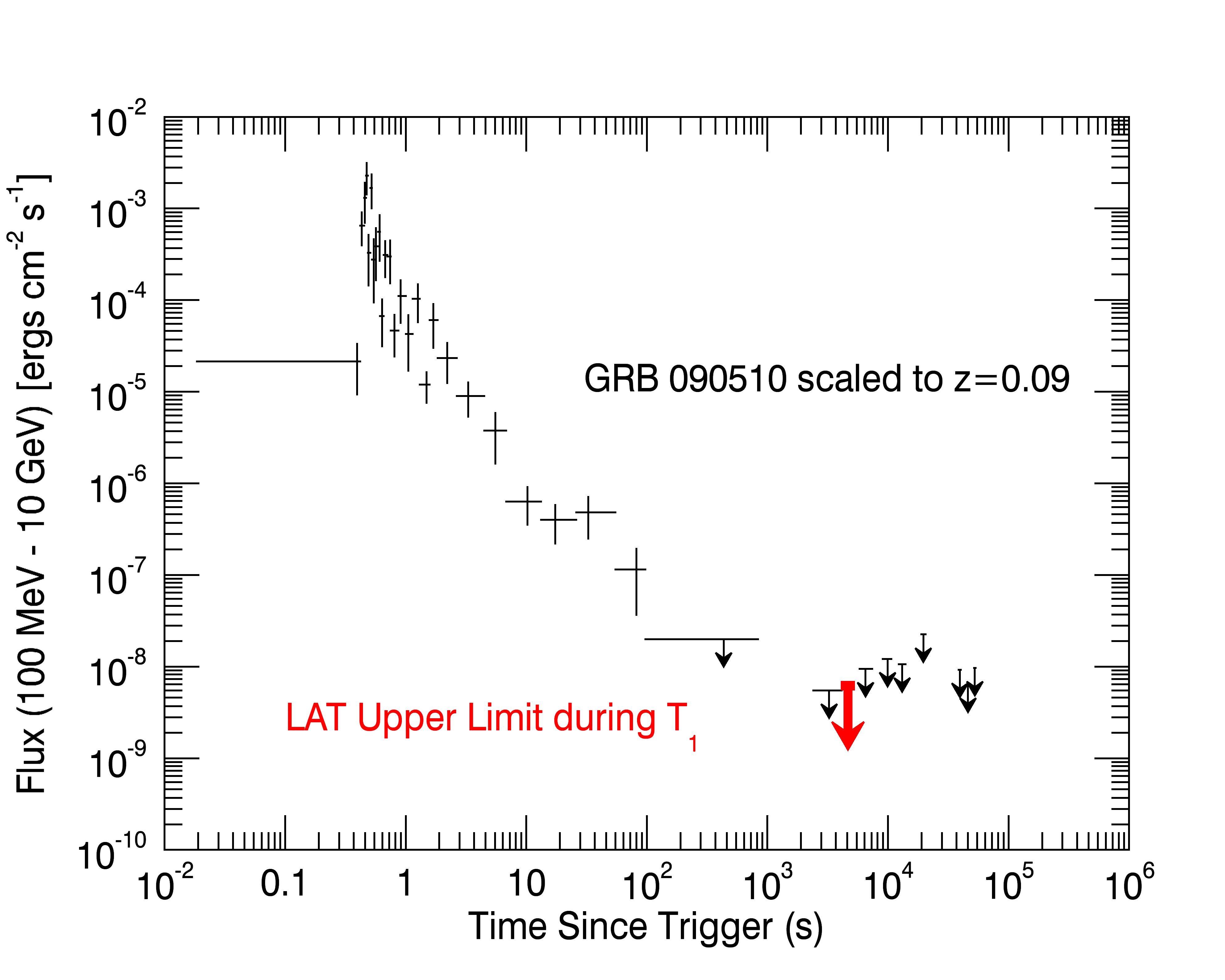}
\caption{
GRB~090510 is the only LAT-detected sGRB with a measured redshift \citep[z = 0.903,][]{2010ApJ...716.1178A}. We compare it here to LAT observations of \GWNAME.  The 100 MeV--10 GeV lightcurve of GRB~090510 has been scaled to z=0.09, the redshift corresponding to the best-fit distance from the GW observation \citep{2041-8205-818-2-L22}. The red arrow indicates the 95\% confidence upper limit measured during $T_1$ across the LIGO localization region and is similar to the upper limits from the measured lightcurve. Even though GRB~090510 had exceptionally bright prompt and afterglow emission, this figure demonstrates that the LAT would detect transients more than an order of magnitude fainter than GRB 090510, provided a more favorable placement of the GW candidate so that it enters the LAT FoV within 100~s of the trigger time.
}
\label{fig:grb090510}
\end{figure}

The most commonly hypothesized EM counterpart of an NS-NS/NS-BH merger is a sGRB, powered by accretion onto one of the two central compact objects 
\citep{1986ApJ...308L..43P,1989Natur.340..126E,1992ApJ...395L..83N,2011ApJ...732L...6R,2012ApJ...746...48M}, 
which launches relativistic jets that produce a short ($<$2 s) bright flash of keV-MeV-peak $\gamma$ rays followed by a broadband longer-lasting afterglow. The LAT detects approximately two sGRBs per year, with $>$100 MeV afterglows lasting up to hundreds of seconds after the trigger \citep{2010ApJ...716.1178A,2013ApJS..209...11A}, seeded by sGRB triggers from the GBM and {\it Swift}-BAT.  The most prominent example of a LAT-detected sGRB is GRB 090510, which simultaneously triggered \Swift and GBM, as well as causing a LAT on-board transient source trigger, resulting in a LAT localization being circulated within seconds. The LAT detected both a spike during the prompt $\gamma$-ray emission and an extended afterglow lasting hundreds of seconds, consistent with the observed optical and X-ray emission \citep{2010ApJ...709L.146D}.  
Figure \ref{fig:grb090510} shows the $>$100 MeV $\gamma$-ray lightcurve of GRB 090510 scaled to z=0.09, the nominal redshift of \GWNAME inferred from the GW observations, in comparison to the LAT upper limits described in \S\ref{sec:short_search}.  If \GWNAME had had a high-energy $\gamma$-ray lightcurve similar to GRB 090510 and had been more favorably placed relative to the LAT boresight at the trigger time, it would have been easily detectable by the LAT during observations similar to those described in this paper. A GRB at the redshift of the LIGO GW source entering the field of view of \Fermi-LAT within the first 100 s would have been detected if it were more than an order of magnitude fainter than GRB 090510.

LAT-detected long GRBs have also shown similar behavior, with a prompt spike contemporaneous with sub-MeV emission followed by long-lived emission lasting from minutes to hours with similar time dependence as radio-to-X-ray afterglows \citep{2013ApJ...763...71A,2014Sci...343...42A}.
Although the detection rate of LAT sGRBs is low, the low redshifts of potentially detectable GW sources and therefore their potentially bright EM counterpart emission, the uncertainty in the observational signatures of GW sources, and the continuous observations of the entire sky during survey operations, make searching the LAT data for counterpart sources worthwhile. 

\subsection{LAT Searches for sGRBs}

Two strategies are useful for associating EM detections of transients with GW sources. The first is to search GW data for counterparts to EM events.  This was done archivally using GRB triggers for the years 2006--2011 \citep{2014PhRvD..89l2004A} and initial LIGO data runs, yielding no credible candidate sources (with a $<20$ Mpc horizon distance). This approach is not very efficient since most of the EM events are outside the current volume sampled by the initial LIGO runs, which increases the trials factors and diminishes the GW sensitivity. 

The second approach is to search for EM bursts (e.g., sGRB) related to GW events.  The most promising way to find a counterpart is via a prompt trigger from the burst itself using wide FoV instruments like GBM or \Swift-BAT.  Given its large FoV and sensitivity to energies from 8 keV to 40 MeV, GBM is the most prolific detector of sGRBs currently operating, with $\sim$40--80 detections per year, and therefore it is very well suited for EM counterpart searches \citep{GBMLIGO}. Yet the LAT offers several capabilities that make it a unique resource. 

In the normal mode of \Fermi GRB detection and measurement, GBM issues triggers on-board for both short and long GRBs (as well as solar flares, terrestrial $\gamma$-ray flashes, magnetar flares, and other short, hard X-ray/soft $\gamma$-ray transients), and approximately half of the GBM GRBs occur in the LAT FoV.  However, in cases of high peak brightness, GBM initiates an automated repoint of the \Fermi spacecraft to optimize LAT observations over the next 2.5 hours. 
Regardless of whether an automated pointing is triggered, even during normal sky survey operations, the LAT will observe the entire sky within 2 orbits ($\sim 3$ hours), fully covering 100\% of any GW localization contour, making follow-up observations of GW triggers both automatic and routine.  Neither instrument on \Fermi has observing constraints limiting the viewing near the Sun or Moon.

If the LAT were to detect a counterpart to a GW trigger, it has the distinct advantage of providing smaller localization uncertainties than GBM.  Typical GBM localizations have radii of several degrees plus systematic uncertainties of 3--14$^\circ$ \citep{2015ApJS..216...32C}, while LAT localizations have radii $\sim 0.1-1^\circ$.  A LAT detection would substantially reduce the sky area for follow-up observations at other wavelengths, some of which require extensive tiling campaigns over days after a trigger to cover a significant fraction of the GW localization region \citep{2016MNRAS.455.1522E,LVEM}.  Although the LAT obtained an on-board localization for short GRB 090510 \citep{2010ApJ...716.1178A} within 15 seconds, LAT localizations typically have an $\sim$8-hr latency for data transmission and ground processing.

The LAT has also recently (as of 2015 June) benefited from a major upgrade to the event reconstruction analysis pipeline, known as Pass 8 \citep{Pass8}.  Pass 8 improves the LAT sensitivity due to increased photon acceptance, especially at low ($<$100 MeV) and high energies ($>$10 GeV), and reduces localization radii by $\sim 30\%$.  Combined with new analysis pipelines that search for transients on all timescales \citep{2015arXiv150203122V}, the LAT is now better suited to discover counterparts to GW sources in both automated pipelines and the specialized searches described in \S\ref{sec:data_analysis}.

The LIGO EM follow-up partners conducted large follow-up campaigns in the optical, radio, and X-ray to search for a counterpart to \GWNAME \citep{LVEM}.  Similar campaigns will occur for other GW triggers no matter the likely progenitor and regardless of whether it is observed along the putative jet axis.
Although no orphan GRB afterglows (which are expected to be associated with off-axis observing angles) have conclusively been detected, models \citep{2002ApJ...570L..61G,2010ApJ...722..235V} predict delayed, fainter transient emission from the afterglow itself, an isotropic optical-near-infrared kilonova signature \citep{1998ApJ...507L..59L}.  Other potential EM counterparts have been theorized, including late radio emission from mildly relativistic material that is dynamically ejected during the merger and drives a shock into the external medium \citep{2011Natur.478...82N}, and radio-to-X-ray (or $\gamma$-ray) emission over seconds to days from relativistic ejecta created as the NSs collide during their merger \citep{2014MNRAS.437L...6K}.
Given the capability of the LAT to detect transients on timescales from milliseconds to years, careful searches for possible afterglow emission within the LIGO localization area over relevant timescales after the GW trigger are essential and require no change to the \Fermi observing strategy.

High-energy $\gamma$-ray emission from GRBs observed by the LAT above 100 MeV lasts much longer (minutes to hours) than the prompt emission observed by the GBM \citep{2013ApJS..209...11A}, for both long and short GRBs \citep{2010ApJ...712..558A,2010ApJ...709L.146D}.  Therefore, even if a localization probability region from LIGO/Virgo is outside the FoV of the LAT during the time of the trigger, the instrument could detect temporally-extended emission in the minutes following the prompt signal, when the region re-enters its FoV. This could result in a much more precise localization of the EM counterpart.  

\subsection{EM Counterparts to BH Binary Mergers}



As discussed in \cite{LIGO_GW150914}, the \GWNAME waveform is consistent with the expectation for the merger of two stellar-mass BHs.  The comparatively clean waveform in the ``chirp'' phase before the merger and the ``ring-down'' stage after the merger would not naturally be expected from coalescing NSs, which possess matter distributed outside an event horizon. The expected progenitors of sGRBs are NS-NS or NS-BH binaries \citep{2013ApJ...762L..18G,1986ApJ...308L..43P,1989Natur.340..126E,1992ApJ...395L..83N,2011ApJ...732L...6R,2012ApJ...746...48M}.  Therefore, a classical sGRB counterpart to \GWNAME is not expected. 

Prior to the discovery of \GWNAME, studies focusing on EM counterparts to stellar-mass BH mergers were few in number.  Most of the numerical simulations have focused on supermassive BH mergers, where circumbinary disk formation is expected with ample gas supply available to power an EM counterpart \citep{2007Sci...316.1874M}.  Stellar-mass BH mergers should require a substantial quantity of nearby gas to form the disk-jet system that is expected to be necessary for an EM counterpart to be detectable at tens-to-hundreds of Mpc distances.  The weak counterpart candidate detected by the GBM described by \cite{GBMLIGO} poses an interesting dilemma for theoretical models, if it is connected to \GWNAME.  Unfortunately, since the GW localization region was not in the FoV of the LAT at the time of the GW trigger, the prompt signal from the GBM candidate could not be addressed with the LAT observations.  Should such $\gamma$-ray
associations arise for future LIGO events, a new paradigm for these mergers will be indicated.

\section{Conclusions} \label{sec:conc}

The \Fermi-LAT is uniquely capable of searching for high-energy $\gamma$-ray counterparts to sources detected by GW observatories.  We use this capability to undertake a detailed search in the regular LAT survey data for a counterpart temporally and spatially coincident with the LIGO trigger on \GWNAME. Although \GWNAME was not in the LAT FoV at the trigger time, the LAT observed the entire region within $\sim 70$ minutes of the GW trigger. 
We searched on short and long timescales for evidence of a transient $\gamma$-ray source contemporaneous with \GWNAME.  No LAT counterpart is detected, and upper limits have been set on the GeV $\gamma$-ray flux within the LIGO localization.   

As the sensitivity of LIGO and Virgo improve over the next few years, their detection horizon for NS-NS and NS-BH binary mergers, and thus the likelihood of sGRB coincidence, will increase greatly.  The merger rates are highly uncertain and depend on the populations and evolution of binaries and opening angles of sGRB jets.  The coincident detection of prompt $\gamma$-ray signals, with on-axis or potentially off-axis afterglow signals will teach us about the physics of binary mergers.
With the discovery of \GWNAME, the search for EM-GW coincidences enters a new phase. It is important to test the strong expectation that stellar BH mergers do not radiate much light; it is equally important to refine the techniques that will be needed to associate sGRBs with GW events from NS mergers. The approach described in this communication is well suited to achieve these twin goals when LIGO detects more GW events.

\acknowledgments{
  The \Fermi LAT Collaboration acknowledges generous ongoing support
from a number of agencies and institutes that have supported both the
development and the operation of the LAT as well as scientific data analysis.
These include the National Aeronautics and Space Administration and the
Department of Energy in the United States, the Commissariat \`a l'Energie Atomique
and the Centre National de la Recherche Scientifique / Institut National de Physique
Nucl\'eaire et de Physique des Particules in France, the Agenzia Spaziale Italiana
and the Istituto Nazionale di Fisica Nucleare in Italy, the Ministry of Education,
Culture, Sports, Science and Technology (MEXT), High Energy Accelerator Research
Organization (KEK) and Japan Aerospace Exploration Agency (JAXA) in Japan, and
the K.~A.~Wallenberg Foundation, the Swedish Research Council and the
Swedish National Space Board in Sweden. Additional support for science analysis during the operations phase is gratefully acknowledged from the Istituto Nazionale di Astrofisica in Italy and the Centre National d'\'Etudes Spatiales in France.
}

\bibliographystyle{yahapj}
\bibliography{references}

\appendix

\section{\Fermi-LAT likelihood analysis}
\label{apex:likelihood}
The standard tools for \Fermi-LAT analysis, the \Fermi \texttt{ScienceTools}, are available for download from the FSSC\footnote{\url{http://fermi.gsfc.nasa.gov/ssc/data/analysis/}}, where the LAT data can also be downloaded. 
In all analyses presented in this paper we used the Pass 8 data class \texttt{P8R2\_TRANSIENT010E\_V6}, and the ST v10r0p5. The \texttt{ScienceTools} are based on the standard maximum likelihood analysis: a model summarizing knowledge about the sources of $\gamma$ rays in a particular region of the sky is convolved with the instrument response and optimized over its parameters to maximize the likelihood that the model best represents the data. The details on how to perform such an analysis with LAT data are described on the FSSC website; here we summarize the main steps. We start by selecting all the data within a given energy range and contained within a ROI. In the case of the \textit{unbinned likelihood} analysis used in this paper, the ROI is circular and is therefore characterized by a center and a radius. Since the upper layers of the Earth's atmosphere are a bright source of $\gamma$ rays that are very difficult to model in the likelihood analysis, we need to further reduce the Earth Limb contamination in the data. 
At the altitude of the nearly-circular \Fermi orbit, the limb is always seen by the LAT at an angle of $\sim 113^\circ$ from the zenith direction. We therefore remove all photons with zenith angles larger than a threshold that depends on the minimum energy used in the analysis (since the PSF is larger at lower energies). In this paper we define the size of the ROI, the energy range and the zenith angle limit used in each analysis in the respective sections. \\

\subsection{Source significance}

The significance of a new source $S$ in a likelihood analysis is determined by using the Likelihood Ratio Test (LRT). The TS of $S$ is equal to twice the logarithm of the ratio of the maximum likelihood value produced with a model including $S$ to the maximum likelihood value of the null hypothesis, i.e., a model that does not include $S$. 
The reference distribution for TS can be used to determine the probability that a measured TS for a source is due to a statistical fluctuation of the null hypothesis (p-value). Unfortunately, the probability density function in a source-over-background test like this cannot, in general, be described by an asymptotic distribution such as expected from Wilks' theorem~\citep{Wilks:38,Protassov2002}.  However, it has been verified by dedicated Monte Carlo simulations~\citep{Mattox96} that the distribution of TS under the null hypothesis is approximately equal to a $\chi^2_{n_{dof}}/2$ distribution\footnote{The factor of \textonehalf ~in front of the TS PDF formula results from allowing only positive source fluxes.}, where $n_{dof}$ is the number of degrees of freedom associated with the new source.  In all cases considered in this paper the new source has a fixed position and a power-law spectrum with two parameters, hence $n_{dof} = 2$. However, in this paper we search on a large region of the sky and over different timescales, and thus we must also account for the trials factor, as explained in the next section.

\subsection{TS maps and trial factors}

Test Statistic maps (TS maps) are used to probe for a source at an unknown location and are obtained by moving a test source by regular steps over a grid and re-optimizing the model parameters to maximize the likelihood. The size of the steps is usually a fraction of the size of the PSF. Since the PSF depends on the energy, we consider the size of the PSF at the minimum energy considered for the map. In our case, we chose a step of 0$\fdg2$. The maximum (denoted TS$_{max}$) of the map corresponds to the most likely localization for a new source. A likelihood analysis with an ROI centered on the position of TS$_{max}$ gives then the final value TS$_{src}$ for the candidate source. However, the search over the grid involves a certain number of non-independent trials, hence the reference distribution of TS$_{src}$ is unknown and must be determined case-by-case with Monte Carlo simulations, as in this work.

\section{TS maps for the long-baseline search}
\label{sec_tsmaps}
For the TS maps of the long-baseline search we use the largest grid fitting inside each of the 9 ROIs. We model the background by taking into account the isotropic component (defined in the template file \texttt{iso\_P8R2\_TRANSIENT010E\_V6\_v06.txt}), which includes the extragalactic diffuse $\gamma$ radiation and residual charged-particle contamination, and the Galactic diffuse $\gamma$-ray emission (using the spatial and spectral template in \texttt{gll\_iem\_v06.fits}), which is the result of interaction of cosmic rays with the gas and the electromagnetic field of the Milky Way. The normalization of the first is left free to vary in order to accommodate orbital variations, while the Galactic diffuse emission model is held fixed to its nominal value. All the point sources from the 3FGL catalog \citep{2015ApJS..218...23A} are also included with their parameters fixed. 

\begin{figure*}[!ht]
\begin{center}
\includegraphics[width=0.32\textwidth]{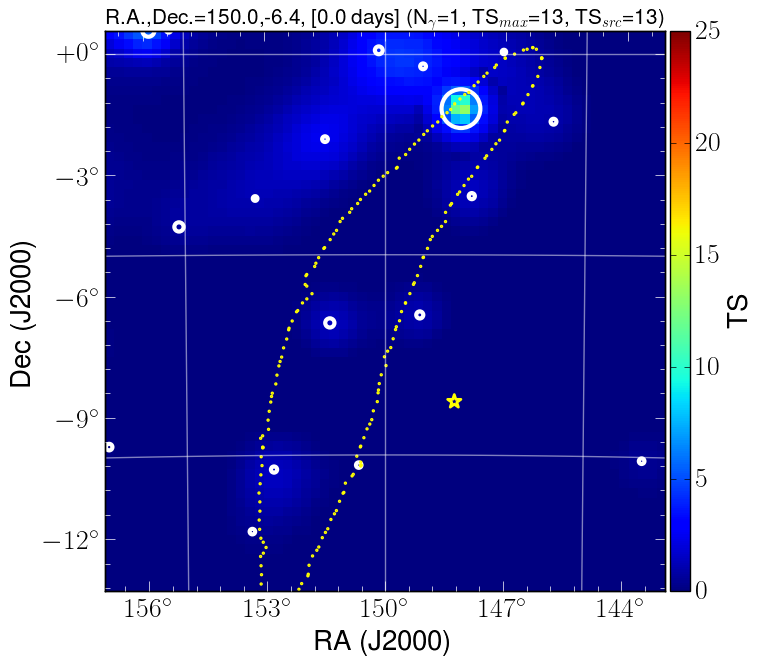}
\includegraphics[width=0.32\textwidth]{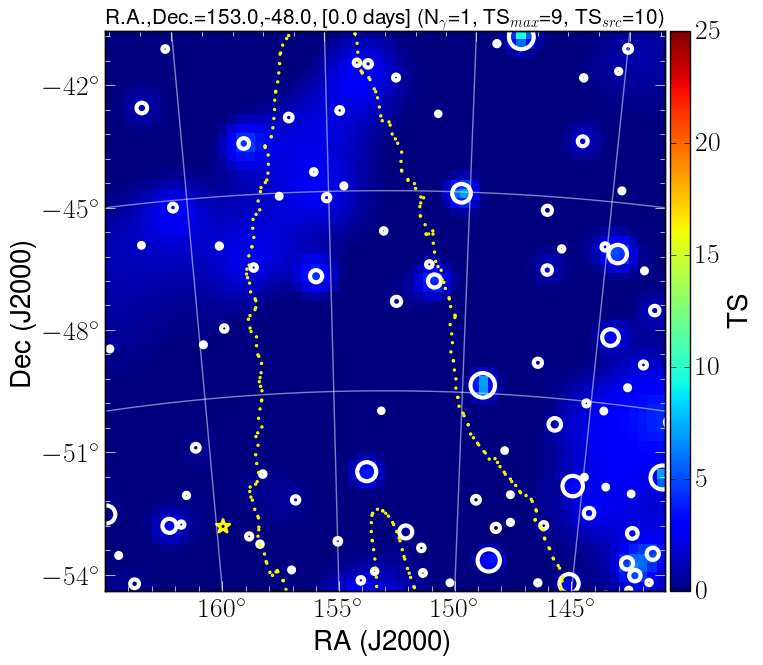}
\includegraphics[width=0.32\textwidth]{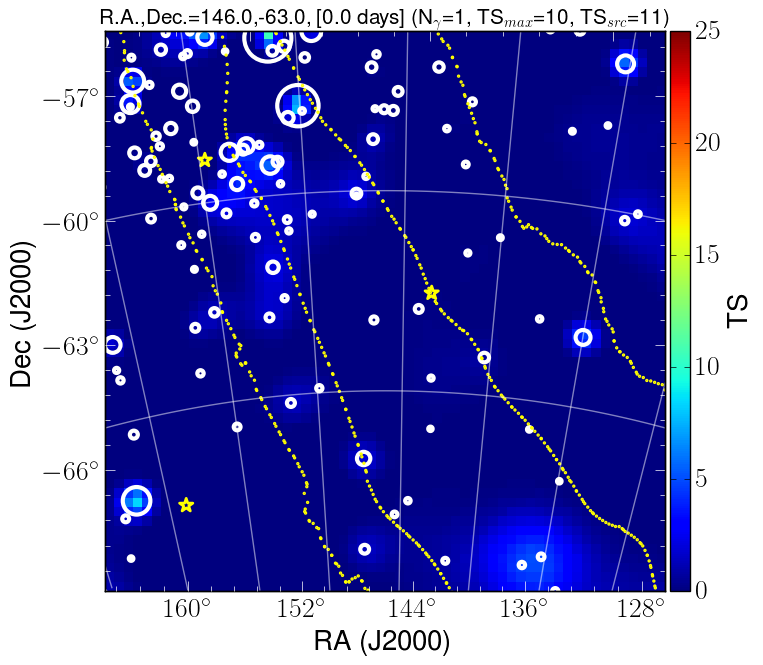}\\
\includegraphics[width=0.32\textwidth]{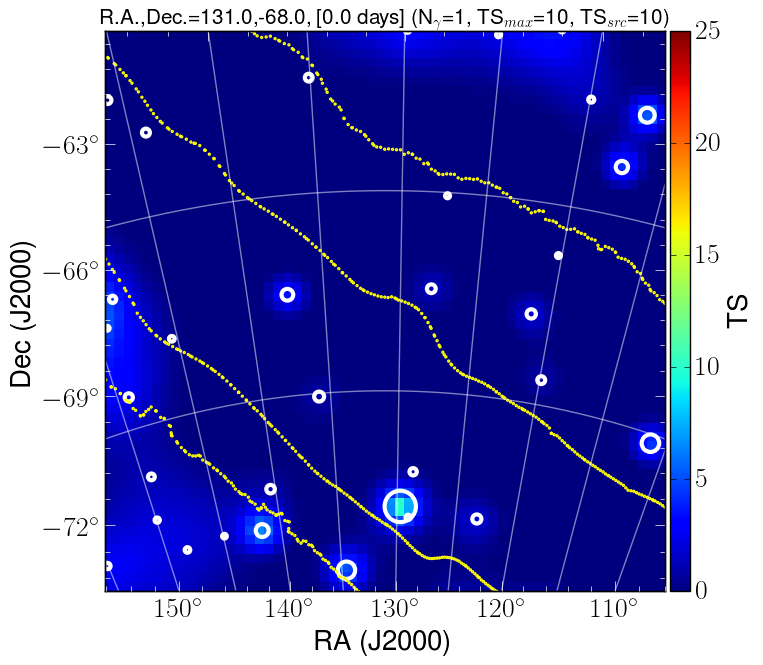}
\includegraphics[width=0.32\textwidth]{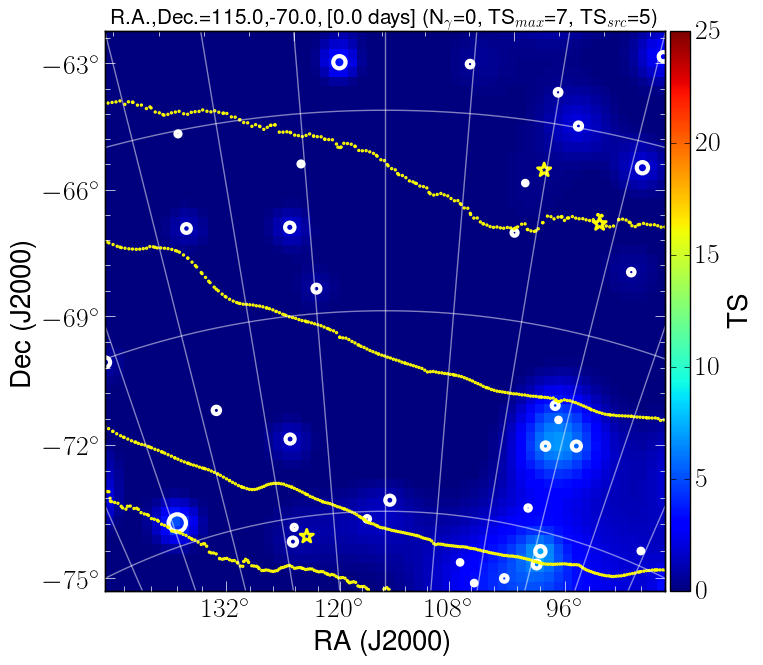}
\includegraphics[width=0.32\textwidth]{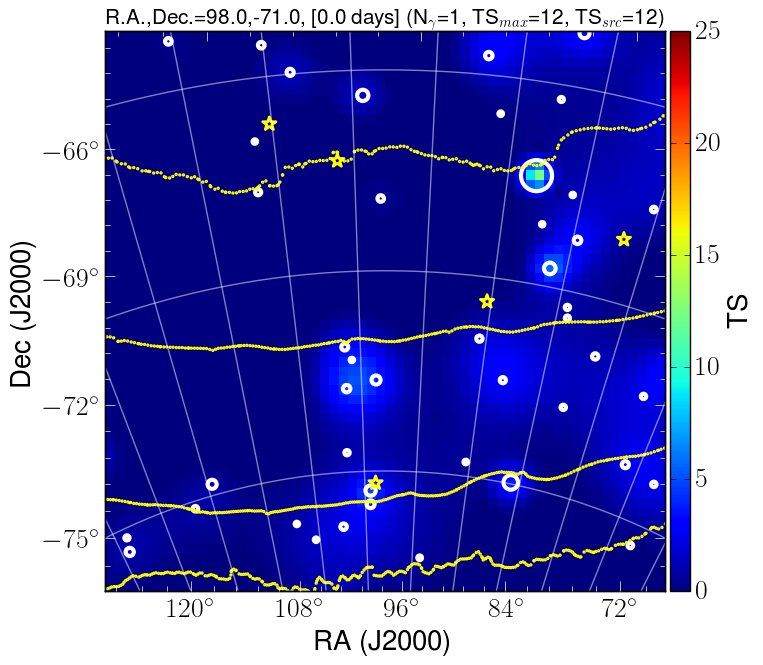}\\
\includegraphics[width=0.32\textwidth]{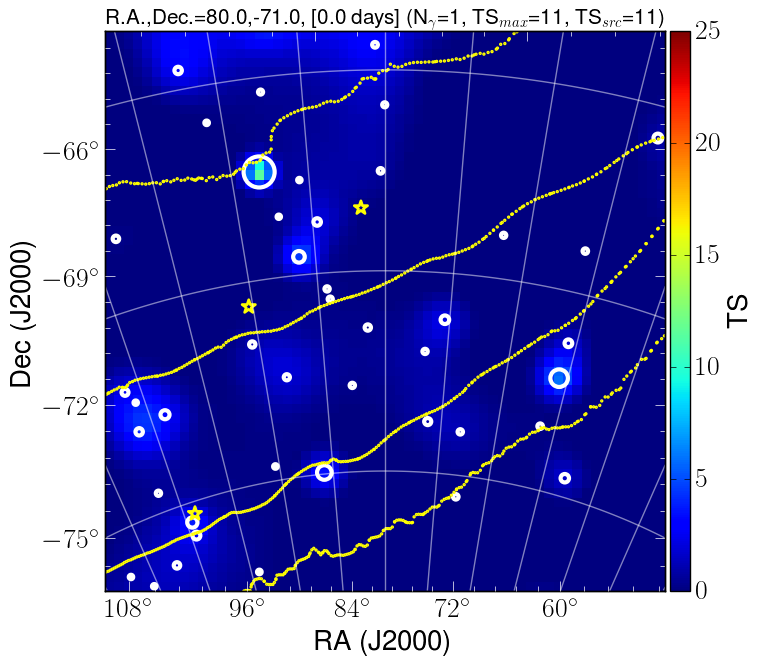}
\includegraphics[width=0.32\textwidth]{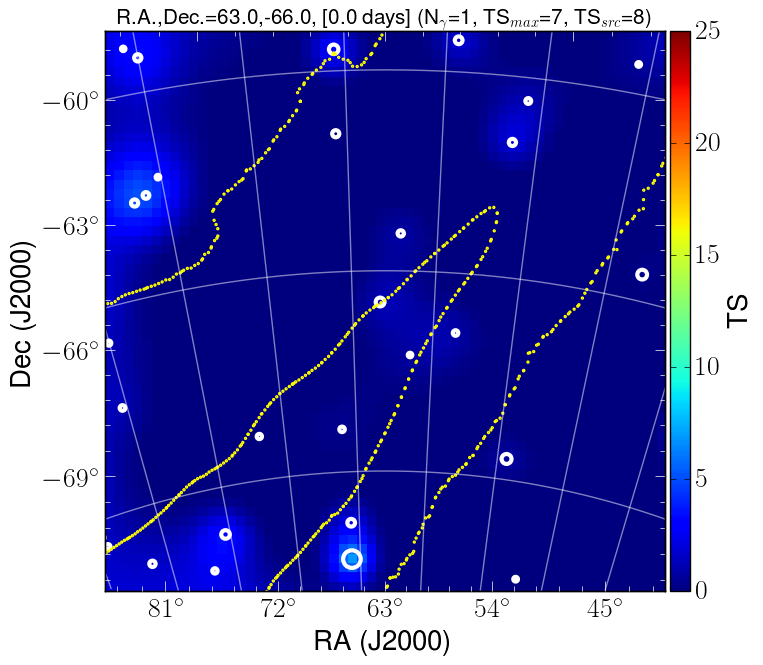}
\includegraphics[width=0.32\textwidth]{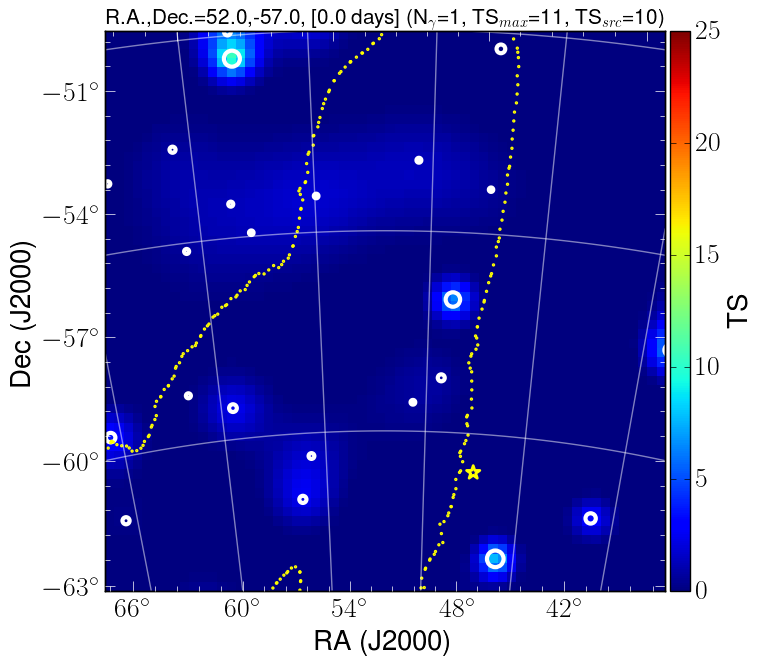}\\
\caption{TS maps for the nine ROIs used in our analysis integrated from $t_{GW}$ to $t_{GW}$+10 ks. Stars indicate the 3FGL sources in each field, the white circles are the individual LAT $\gamma$ rays (with the size proportional to their reconstructed energy). The yellow dots trace the LIGO 90\% contour.
}
\label{fig:tsmap1d}
\end{center}
\end{figure*}

In Figure \ref{fig:tsmap1d} we display the TS maps for each region integrated from $t_{GW}$ to $t_{GW}$+10 ks. 
The 3FGL sources have been overlaid as yellow stars while the white circles represent the \Fermi-LAT $\gamma$ rays (the size of the circle is proportional to the reconstructed energy of the event).

\section{\Fermi-LAT point sources}
\label{sec_3fgsource}

For completeness, we report the \Fermi-LAT sources in the 3FGL catalog that are present in each ROI, with, where available, source and class associations. Since the integration times in our analysis are much shorter than the integration time of the catalog, we select only those sources with an average significance (\texttt{Signif\_Avg} in the 3FGL) $>$20, which roughly corresponds to an average flux in the range 10$^{-9}$--10$^{-7}$ ph\,cm$^{-2}$\,s$^{-1}$ in the energy range 0.1--1 GeV.
\begin{deluxetable}{l c c l l l}  
\tablecolumns{6}
\tablecaption{ROIs and the 3FGL sources they contain}
\tablehead{   
  \colhead{ROI (R.A.,Dec.)} &
  \colhead{R.A.} &
  \colhead{Dec.} &
  \colhead{3FGL Name} &
  \colhead{Association} &
  \colhead{Class\tablenotemark{a}}\\
  & \colhead{deg} & \colhead{deg} & & &
  }
\startdata
ROI$_{0}$  (150.0, $-$6.4)  & 147.223 & 0.363 & 3FGL J0948.8+0021 & PMN J0948+0022 & NLSY1 \\ 
                           & 148.255 & $-$8.663 & 3FGL J0953.0-0839 & PMN J0953-0840 & bll \\ 
\hline
ROI$_{1}$  (153.0, $-$48.0)  & 155.790 & $-$57.759 & 3FGL J1023.1-5745 & LAT PSR J1023-5746 & PSR \\ 
                           & 159.735 & $-$53.186 & 3FGL J1038.9-5311 & MRC 1036-529 & bcu \\ 
                           & 164.499 & $-$52.455 & 3FGL J1057.9-5227 & PSR J1057-5226 & PSR \\ 
\hline
ROI$_{2}$  (146.0, $-$63.0)  & 136.218 & $-$57.570 & 3FGL J0904.8-5734 & PKS 0903-57 & bcu \\ 
                           & 143.481 & $-$62.534 & 3FGL J0933.9-6232 &  &  \\ 
                           & 154.730 & $-$58.946 & 3FGL J1018.9-5856 & 1FGL J1018.6-5856 & HMB \\ 
                           & 155.790 & $-$57.759 & 3FGL J1023.1-5745 & LAT PSR J1023-5746 & PSR \\ 
                           & 157.123 & $-$58.320 & 3FGL J1028.4-5819 & PSR J1028-5819 & PSR \\ 
                           & 158.926 & $-$67.334 & 3FGL J1035.7-6720 &  &  \\ 
                           & 161.129 & $-$57.630 & 3FGL J1044.5-5737 & LAT PSR J1044-5737 & PSR \\ 
                           & 161.277 & $-$59.692 & 3FGL J1045.1-5941 & Eta Carinae & BIN \\ 
                           & 162.067 & $-$58.535 & 3FGL J1048.2-5832 & PSR J1048-5832 & PSR \\ 
\hline
ROI$_{3}$  (131.0, $-$68.0)  & 122.811 & $-$75.492 & 3FGL J0811.2-7529 & PMN J0810-7530 & bll \\ 
                           & 143.481 & $-$62.534 & 3FGL J0933.9-6232 &  &  \\ 
\hline
ROI$_{4}$  (115.0, $-$70.0)  & 90.313 & $-$70.609 & 3FGL J0601.2-7036 & PKS 0601-70 & fsrq \\ 
                           & 98.942 & $-$75.293 & 3FGL J0635.7-7517 & PKS 0637-75 & fsrq \\ 
                           & 101.088 & $-$67.223 & 3FGL J0644.3-6713 & PKS 0644-671 & bcu \\ 
                           & 105.158 & $-$66.173 & 3FGL J0700.6-6610 & PKS 0700-661 & bll \\ 
                           & 122.811 & $-$75.492 & 3FGL J0811.2-7529 & PMN J0810-7530 & bll \\ 
\hline
ROI$_{5}$  (98.0, $-$71.0)  & 81.650 & $-$68.420 & 3FGL J0526.6-6825e & LMC & GAL \\ 
                           & 90.313 & $-$70.609 & 3FGL J0601.2-7036 & PKS 0601-70 & fsrq \\ 
                           & 98.942 & $-$75.293 & 3FGL J0635.7-7517 & PKS 0637-75 & fsrq \\ 
                           & 101.088 & $-$67.223 & 3FGL J0644.3-6713 & PKS 0644-671 & bcu \\ 
                           & 105.158 & $-$66.173 & 3FGL J0700.6-6610 & PKS 0700-661 & bll \\ 
                           & 122.811 & $-$75.492 & 3FGL J0811.2-7529 & PMN J0810-7530 & bll \\ 
\hline
ROI$_{6}$  (80.0, $-$71.0)  & 79.189 & $-$62.121 & 3FGL J0516.7-6207 & PKS 0516-621 & bll \\ 
                           & 81.650 & $-$68.420 & 3FGL J0526.6-6825e & LMC & GAL \\ 
                           & 90.313 & $-$70.609 & 3FGL J0601.2-7036 & PKS 0601-70 & fsrq \\ 
                           & 98.942 & $-$75.293 & 3FGL J0635.7-7517 & PKS 0637-75 & fsrq \\ 
                           & 101.088 & $-$67.223 & 3FGL J0644.3-6713 & PKS 0644-671 & bcu \\ 
\hline
ROI$_{7}$  (63.0, $-$66.0)  & 47.478 & $-$60.963 & 3FGL J0309.9-6057 & PKS 0308-611 & fsrq \\ 
                           & 76.780 & $-$61.050 & 3FGL J0507.1-6102 & PKS 0506-61 & fsrq \\ 
                           & 79.189 & $-$62.121 & 3FGL J0516.7-6207 & PKS 0516-621 & bll \\ 
                           & 81.650 & $-$68.420 & 3FGL J0526.6-6825e & LMC & GAL \\ 
\hline
ROI$_{8}$  (52.0, $-$57.0)  & 39.195 & $-$61.600 & 3FGL J0236.7-6136 & PKS 0235-618 & fsrq \\ 
                           & 47.478 & $-$60.963 & 3FGL J0309.9-6057 & PKS 0308-611 & fsrq
\enddata
\tablenotetext{a}{Class names are from the 3 FGL catalog \citep{2015ApJS..218...23A}.}
\label{table:roi}
\end{deluxetable}

\end{document}